\documentclass[aps,prl,twocolumn,showpacs,preprintnumbers,amsmath,amssymb,floatfix]{revtex4-1}
\usepackage{graphicx}
\usepackage{dcolumn}
\usepackage{bm}
\usepackage{color}
\usepackage{epstopdf}%
\usepackage{amsmath}%
\usepackage{amsfonts}%
\usepackage{amssymb}
\providecommand{\U}[1]{\protect\rule{.1in}{.1in}}

\newcommand{\be}{\begin{equation}}
\newcommand{\ee}{\end{equation}}
\newcommand{\bea}{\begin{eqnarray}}
\newcommand{\eea}{\end{eqnarray}}

\newcommand{\mb}{\mathbf}

\newcommand{\mbf}{\mathbf}

\begin{document}

\title{Closed-form weak localization magnetoconductivity in quantum wells with arbitrary Rashba and Dresselhaus spin-orbit interactions}
\author{D. C. Marinescu}
\affiliation{Department of Physics and Astronomy, Clemson University,Clemson, South Carolina 29634, USA}
\author{Pirmin J. Weigele}
\affiliation{{Department of Physics,
University of Basel,
Klingelbergstrasse 82,
CH-4056 Basel, Switzerland}}
\author{J. Carlos Egues}
\affiliation
{Instituto de F\'isica de S\~ao Carlos, Universidade de S\~ao Paulo,
13560-970 S\~ao Carlos, S\~ao Paulo, Brazil}
\author{Dominik M. Zumb\"{u}hl}
\affiliation{{Department of Physics,
University of Basel,
Klingelbergstrasse 82,
CH-4056 Basel, Switzerland}}

\begin{abstract}
We derive a closed-form expression for the weak localization (WL) corrections to the
magnetoconductivity of a 2D electron system with arbitrary Rashba $\alpha$ and Dresselhaus $\beta$ (linear) and $\beta_3$ (cubic)
spin-orbit interaction couplings, in a perpendicular magnetic field geometry. In a system of reference with an in-plane $\hat{z}$ axis chosen as the high spin-symmetry direction at $\alpha = \beta$, we formulate a new algorithm to calculate the three independent contributions that lead to WL. The antilocalization is counterbalanced by the term associated with the spin-relaxation along $\hat{z}$, dependent only on $\alpha - \beta$. The other term is generated by two identical scattering modes characterized by spin-relaxation rates which are explicit functions of the orientation of the scattered momentum. Excellent agreement is found with data from GaAs quantum wells, where in particular our theory correctly captures the shift of the minima of the WL curves as a function of $\alpha/\beta$. This suggests that the anisotropy of the effective spin relaxation rates is fundamental to understanding the effect of the SO coupling in transport.
\end{abstract}
\date{\today}
\maketitle

\textbf{Introduction -} The sensitivity of transport measurements to the weak-localization (WL) phenomenon in spin-orbit (SO) coupled systems placed in a perpendicular magnetic field has stimulated, for over twenty years, a continuous theoretical effort in providing models that can lead to the extraction of the SO parameters from transport data. Of interest are linear Rashba \cite{rashba} $\alpha$ coupling, which results from the broken inversion symmetry in III-V semiconductor quantum wells, and the linear $\beta_1$ and cubic $\beta_3$ Dresselhaus \cite{dresselhaus} couplings which originate in the broken inversion symmetry in the crystal structure.

Ideally, for convenience in reliably fitting experimental data, the theoretical result should be a closed-form expression. Thus far, since the publication of the Larkin-Hikami-Nagaoka formula \cite{Hikami1980} in 1980 for a non-specific case of a spin-orbit interaction, obtaining analytical results in semiconductor systems has been limited to two situations: when only one linear coupling, either linear Rashba \cite{Punnoose2006,ketterman} or the renormalized linear Dresselhaus term $\beta = \beta_1-\beta_3$ \cite{iordanskii} is present, or when $\alpha$ and $\beta$ are exactly equal \cite{pikus1,kammermeier}. In the transitory regime to the $\alpha = \beta$ state, a closed form expression that includes both the linear terms was obtained only recently in \cite{prx}. In general, the simultaneous inclusion in the WL problem of both linear terms, which rotate the spin in opposite directions, was manageable to date only numerically \cite{pikus1,pikus2, golub-prb,golub-2006,golub-2009,kohda12}.

Computationally, evaluating the WL correction to magnetoconductivity requires an integration in the momentum space of the eigenvalues of the Cooperon, an operator that describes the renormalization of the impurity scattering matrix element on account of the interference of the incident and back-scattered spin $1/2$ electron states. Because the spin-states are mixed by the SO interaction, the Cooperon acts in the tensor product quantum space between a position representation and the four dimensional space associated with the total spin angular momentum $\mathbf{J} = (J_x, J_y, J_z)$ of eigenvalues $J = 0$ and $J=1$.

In this paper we formulate a general decoupling algorithm for the triplet Cooperon modes that leads to a representation invariant, closed-form expression for the magnetoconductivity valid for any values of $\alpha$, $\beta$, and $\beta_3$. Essential to this approach is the recognition that the spin-relaxation rates can be dependent on the angular orientation of the scattering momentum. Three different scattering processes are found to contribute to the WL corrections.

The $|1,0\rangle_z$ mode associated with the in-plane direction that becomes the high spin-symmetry axis when $\alpha = \beta$, depends only on the difference $\alpha - \beta$ and $\beta_3$, and in the limit $\alpha \rightarrow \beta, \beta_3\rightarrow 0$ exactly cancels the antilocalization correction, $|0,0\rangle_z$. The $\beta_3\neq 0$ coupling determines the difference between these terms in the high spin-symmetry state $\alpha = \beta$. The other two modes, superpositions of $|1,0\rangle_x$ and $|1,0\rangle_y$, are characterized by the same effective spin-relaxation rate that is dependent on the angular orientation of the scattering momentum. Their logarithmic contributions to the WL are averaged over the angular distribution in the final expression. Our theory is able to reproduce remarkably well quantitatively and qualitatively the experimental features detected in Ref.~\cite{prx}, especially the trademark of the SO coupling, the reduction of the magnetic field value at which the minimum magnetoconductivity is obtained when $\alpha/\beta$ increases from zero to one.

\textbf{The Weak Localization Formalism -} The 2D electron system with linear Rashba $\alpha >0 $, linear $\beta >0$ and cubic $\beta_3$ Dresselhaus SO interactions is described in a rotated $\hat{x}-\hat{z}$ reference frame with $x\,\parallel [110]$ and $z\,\parallel [1\bar{1}0]$ \cite{mar3, prx}. This choice of axes permits a symmetric description of the SO couplings and highlights the existence of a privileged in-plane direction which we choose as the quantization axis of the $z$ component of the electron spin when the linear interactions that rotate the spin in opposite directions match. When $\alpha < 0$ this system of coordinates is rotated by $\pi/2$, such that when $\alpha + \beta = 0$, the spin quantization occurs along the new $\hat{z}$ axis.

The single-particle
Hamiltonian of a conduction electron of effective mass $m^*$, Fermi momentum $\mathbf{p} = \hbar k_F(\cos\varphi_{\mb p}, \sin\varphi_{\mb p})$,
and spin $\mathbf{\sigma }=(\sigma _{x},\sigma _{y},\sigma _{z})$ can be written in a symmetrized form in the SO components, as \cite{mar3,prx},
\begin{equation}
H_{\mb p} = \frac{p^{2}}{2m^*} + \hbar\left( \mathbf{\Omega}_{\mb p}\times\mathbf{\sigma}\right)
\cdot\hat{y}\;,\label{eq:delta-h}
\end{equation}
where $\hbar\Omega_{\mb p}$ is the effective spin-orbit field with components,
\begin{align}
\hbar\Omega_{\mb p}^x  &  =  k_F\left[(\alpha + \beta)\cos\varphi_{\mb p}- \beta_{3}\cos
3\varphi_{\mb p}\right]\;,\nonumber\\
\hbar\Omega_{\mb p}^z  &  =  k_F\left[(\alpha - \beta)\sin\varphi_{\mb p}- \beta_{3}\sin
3\varphi_{\mb p}\right]\;.
\end{align}

In the following considerations, we assume that scattering of impurities is elastic, spin-independent, and involves only states at the Fermi surface, whose density per unit area, for a single spin is $\nu_{0} = m^{*}/2\pi\hbar^{2}$. The scattering matrix element of two electrons of momenta $\mbf p$ and $\mbf p'$, $|V_{\mbf p,\mbf p'}|^2$, depends only on the angle between the incident and scattered directions.

The quantum corrections to the conductivity are calculated from the general
expression \cite{gorkov},
\begin{equation}
\Delta\sigma = -\frac{2e^{2}D}{\hbar}\sum_{\mathbf{q}, i}%
\frac{1}{\cal E}_{i}(\mathbf{q})\;,\label{eq:cond-gen}
\end{equation}
where $D = v^2\tau_1/2$ is the diffusion coefficient in two dimensions expressed as a function of the transport time, $\tau_1$. This is the first $(n =1)$ in a series of transport times that result from the anisotropy of the scattering matrix element \cite{pikus2}
\begin{equation}
\frac{\hbar}{\tau_{n}} = \nu_{0}\int
|V_{\mathbf p,\mathbf p'}|^2(1-\cos n\varphi)d\varphi\;,\label{eq:n-times}
\end{equation}
${\cal{E}}_{i}(\mathbf{q})$ ($ i =\overline{0,3}$) are the eigenvalues of the Cooperon operator which renormalizes the scattering matrix element upon impurity averaging for an electron state $\mb p$ that is almost perfectly backscattered into $\mb p' \approx -\mb p$. The deviation from this situation is represented by $\mb q = \mb p + \mb p'$, with $q\ll p$.

Following the weak localization formalism, the Cooperon operator is found to be \cite{mar3, prx},
\begin{eqnarray}
\mathcal{H} & = & Dq^{2} + \frac{1}{\tau_{\phi}} + D\left\{ \left[
Q_{S}^{2} + Q_{3}^{2}\right] J_{z}^{2} + \left[ Q_{A}^{2} + Q_{3}^{2}\right]
J_{x}^{2} \right. \nonumber\\
& + &\left.  2 Q_{A}q_{z}J_{x}-2Q_{S}q_{x}J_{z}\right\} \;, \label{eq:calh}
\end{eqnarray}
where the SO couplings are encapsulated in the equivalent momenta,
\begin{equation}
Q_{S}  =  \frac{2m^*(\alpha + \beta)}{\hbar^2} \;,\;
Q_{A}  =  \frac{2m^*(\alpha - \beta)}{\hbar^2}\;,\;
Q_{3}  = \frac{2m^*\beta_3}{\hbar^2}\sqrt{\frac{\tau_{3}}{\tau_{1}}}\;. \label{eq:q}
\end{equation}
$\tau_\phi$ is the dephasing time, a measure of the inelasticity of the scattering process.

\textbf{The Cooperon Equation -}  We diagonalize the Cooperon ${\cal H}$ in a basis formed by the following eigenstates of the angular momentum $\mathbf{J}^2$:$\{|0,0\rangle_z,|1,0\rangle_z, |10\rangle_x, |1,0\rangle_y\}$ where $|10\rangle_x = (|11\rangle_z-|1-1\rangle_z)/\sqrt{2}$ and $|10\rangle_y =  (|11\rangle_z+|1-1\rangle_z)/\sqrt{2}$ (the subindex indicates the corresponding projection of the total angular momentum). The singlet contribution, $|0,0\rangle_z$, is separable on account of the orthogonality of the $\mb J^2$ eigenstates and has an eigenvalue,
\begin{equation}
{\mathcal{E}}_{0} = Dq^2 + \frac{1}{\tau_\phi}\;.\label{eq:singlet}
\end{equation}
Its contribution to WL is negative, the antilocalization, because of the antisymmetric character of the spins state under the spin permutation.

Independently, the triplet modes satisfy a characteristic equation,
\begin{align}
&(E_1-\mathcal{E})(E_2-\mathcal{E})(E_3-\mathcal{E}) \nonumber\\
&-4D^2Q_A^2q_z^2(E_3-\mathcal{E}) - 4D^2Q_S^2q_x^2(E_2-\mathcal{E}) = 0\;,\label{eq:cubic-final}
\end{align}
with
\bea
E_1&=& Dq^2 + \frac{1}{\tau_\phi} +D\left({Q}_S^2 + {Q}_A^2+2Q_3^2\right)\;,\nonumber\\
E_2& = & Dq^2 + \frac{1}{\tau_\phi}+ D({Q}_A^2 + Q_3^2)\;,\nonumber\\
E_3 & = & Dq^2 + \frac{1}{\tau_\phi}+ D({Q}_S^2 + Q_3^2)\;.
\eea

In a first order approximation that neglects any off-diagonal contributions, the Cooperon eigenvalues are ${\mathcal{E}_{i}} = E_i$, a result, which upon integration over $\mb{q}$ leads to the generalization of the $\alpha  = 0$ case in Ref.~\onlinecite{iordanskii} for $B = 0$.

An improvement on this approximation is obtained by calculating directly the sum of the inverse eigenvalues from the coefficients of Eq.~(\ref{eq:cubic-final}). Adding the singlet contribution an {\it exact} solution for the conductivity correction is obtained in terms of quadratures from Eq.~(\ref{eq:cond-gen}), as
\begin{align}
&\Delta \sigma(0) = -\frac{De^2}{2\pi^2\hbar}\int_{q_{min}}^{q_{max}}qdq\int_0^{2\pi}d\varphi \left[-\frac{1}{\mathcal{E}_{0}} \right.\nonumber\\
&\left.+ \frac{E_1E_2+E_2E_3+E_3E_1 - 4D^2q^2 \left(Q_A^2\sin^2\varphi + Q_S^2\cos^2\varphi\right)}
{E_1E_2E_3-4D^2\left(E_3 Q_A^2\sin^2\varphi + E_2 Q_S^2\cos^2\varphi\right)}\right]\;,\label{eq:coop-sum}
\end{align}
where the limits of integration are $q_{\min} = 0$ and $q_{\max} = 1/\sqrt{D\tau_1}$. Characteristic to this expression is the coupling between the triplet modes that is realized by the SO off-diagonal terms.

The calculation of the magneto-conductivity follows a similar path, since it is based on applying a unitary transformation generated by the magnetic vector potential
$\mathbf{A }= (A_{x} = Bz, A_{y} = 0, A_{z} = 0)$ on Eq.~(\ref{eq:calh})  \cite{altshuler, prx}.
In this algorithm, the structure of the Cooperon equation in spin space is preserved, while the orbital motion is quantized in Landau levels of index $n$ with $q^{2}$ being replaced by $\frac{4eB}{\hbar}\left(n + \frac{1}{2}\right)$ \cite{altshuler}.

If in Eq.~(\ref{eq:cubic-final}) we assume that solutions $\mathcal{E}$ depend only on $q^2$,  upon angular averaging, $\langle q_x^2\rangle = \langle q_z^2\rangle = q^2/2 = \frac{2eB}{\hbar}(n+1/2)$, and the magnetoconductivity can be calculated from Eq.~(\ref{eq:coop-sum}), modified into a sum over all LLs. The ensuing result is a representation-invariant generalization of the magnetoconductivity obtained in Ref.~\cite{iordanskii} for $\alpha = 0$.

As a function of the magnetic field, however, the magnetoconductivity calculated in this way reaches a minimum at the same $B$ for all $\alpha/\beta$ values, thus unable to reproduce the displacement of $B_{\min}$ seen experimentally \cite{kohda12,prx-1}. This discrepancy suggests that the observed magnetoconductivity correction cannot be generated by coupled triplet scattering modes and a different approach is needed to find the eigenvalues of the Cooperon.

\textbf{Decoupled Cooperon Modes -} We solve Eq.~(\ref{eq:cubic-final}) by proposing the following form for its solutions,
\be
{\mathcal E} = Dq^2 + \frac{1}{\tau_\phi} + 2Dq_0(\varphi) q + \mu(\varphi)\;,
\ee
where $q_0$ and $\mu$ are functions of the angle $\varphi$ made by $\mb{q}$ with the $\hat{x}$ axis.
With, the Dyakonov-Perel spin relaxation rates along the $x, z, y$ axes respectively,
\begin{eqnarray}
& & \xi_3 = D\left(Q_{S}^{2} + Q_3^2\right)\;,\nonumber\\
& & \xi_2 = D\left(Q_{A}^2 + Q_3^2\right)\;,\nonumber\\
& & \xi_1 = D\left(Q_{A}^{2} + Q_S^2 + 2Q_3^2\right) \;,\label{eq:spin-relaxation}
\end{eqnarray}
we denote $\varepsilon_i = \xi_i -\mu$.

We treat Eq.~(\ref{eq:cubic-final}) as a polynomial identity in $q$ and equate the coefficients of all the independent terms to obtain the following system of equations:
\begin{subequations}\label{eq:nu-U}
\begin{align}
& \varepsilon_1\varepsilon_2\varepsilon_3 = 0;\label{eq:zero}\\
& q_0(\varepsilon_1\varepsilon_2 + \varepsilon_2\varepsilon_3 + \varepsilon_3\varepsilon_1) = 0\;,\label{eq:unu}\\
& q_0^2(\varepsilon_1+\varepsilon_2+\varepsilon_3) = Q_S^2\varepsilon_2\cos^2\varphi + Q_A^2\varepsilon_3\sin^2\varphi\;,\label{eq:doi}\\
& q_0^3 = q_0(Q_A^2\sin^2\varphi + Q_S^2\cos^2\varphi)\;.\label{eq:trei}
\end{align}
\end{subequations}
Equation (\ref{eq:nu-U}) always admits the trivial solution $q_0 = 0$, $\varepsilon_i = 0$, or equivalently $\mu_i = \xi_i$, the diagonal matrix elements in Eq.~(\ref{eq:cubic-final}).

Here we demonstrate that a set of non-trivial solutions is also available.
Eq.~(\ref{eq:trei}) actually generates three solutions for $q_0$, each associated, respectively, with the three Cooperon eigenvalues we are looking for:
\begin{equation}
q_{0} = 0\;,\quad \pm \sqrt{Q_A^2\sin^2\varphi + Q_S^2\cos^2\varphi}\;.\label{eq:q0}
\end{equation}
For $q_0 = 0$, Eq.~(\ref{eq:zero}) has to be satisfied independently by setting $\varepsilon_2 = 0$. This choice, which leads to $\mu_2 = \xi_2$, is supported by the form of the solution of Eq.~(\ref{eq:cubic-final}) in the two limiting cases $Q_A = Q_S$ and $Q_A = 0$ when it can be solved exactly. ($\varepsilon_3 = 0$, leading to $\mu_3 = \xi_3$, is the appropriate solution for $Q_S = 0$.) The corresponding Cooperon eigenvalue is therefore,
\be
\mathcal{E}_2 = Dq^2 + \frac{1}{\tau_\phi} + D(Q_A^2 + Q_3)^2\;, \label{eq:mode2}
\ee
a scattering mode associated with the $|1,0\rangle_z$ eigenstate. Both the antilocalization contribution Eq.~(\ref{eq:singlet}), determined by $|0,0\rangle_z$ state, and ${\mathcal{E}}_2$ describe dispersions localized at the origin in the momentum $\mb q$ space.

 When $\varepsilon_2 = 0$, Eq.~(\ref{eq:doi}) is satisfied for as long as terms of order four in SO constants are neglected since $Q_A^2\varepsilon_3 = D(Q_S^2-Q_A^2)Q_A^2$. Equation(\ref{eq:doi}) is satisfied exactly when $Q_A = 0$ or when $Q_A = Q_S$.

When inserted in Eq.~(\ref{eq:doi}), the other two solutions for $q_0$ generate the same value of $\mu$ for the remaining Cooperon modes,
\be
\mu_{1,3}(\varphi) = \frac{(\xi_1 + \xi_3)Q_S^2\cos^2\varphi + (\xi_1 + \xi_2)Q_A^2\sin^2\varphi}{2(Q_S^2\cos^2 \varphi + Q_A^2\sin^2\varphi)}\;.
\ee
From Eq.~(\ref{eq:q0}) we define an in-plane vector $\mb{q}_0 = (Q_S\cos\varphi, Q_A\sin\varphi)$,
which makes an angle $\theta$ with $\mb{q}$ given by,
$
\cos \theta (\varphi) = \left(Q_S\cos^2\varphi + Q_A\sin^2\varphi\right)/\sqrt{Q_S^2\cos^2\varphi + Q_A^2\sin^2\varphi}$.
With this, the solutions $1,3$ of the Cooperon equation become,
\begin{equation}
{\mathcal {E}}_{1,3} = D(\mb{q}\pm\mb{q}_0/\cos\theta)^2 + \frac{1}{\tau_\phi} + \mu_{1,3}(\varphi) -\frac{q_0^2(\varphi)}{\cos^2\theta(\varphi)}\;,\label{eq:dispersion}
\end{equation}
describing two independent dispersions separated in the momentum space by $2\mb{q}_0(\varphi)/\cos \theta(\varphi)$, associated with linear combinations of $|1,0\rangle_x$ and $|1,0\rangle_y$ spin states, of identical effective anisotropic spin-relaxation  rate,
\begin{equation}
\frac{1}{\tau_{eff}(\varphi)} = \mu_{1,3}(\varphi) -\frac{q_0^2}{\cos^2\theta(\varphi)}\;.\label{eq:taueff}
\end{equation}

The exploration of the anisotropic character of the SO interaction and its effect on the electron spin relaxation rates, is not possible in the usual full quantization formalism, in which the operator expressions of $q_x$ and $q_z$ do not commute \cite{iordanskii, altshuler,prx}. Presently, the only representation-invariant solutions to the WL problem in SO coupled systems obtained in this way correspond to the $\alpha = \beta$ limit \cite{prx}. In all other cases, the WL magnetoconductivity depends explicitly on off-diagonal SO matrix elements, so a simple change of axes leads to a different numerical result \cite{iordanskii, golub-prb, golub-2006,golub-2009}. Here, the quantization of the electron orbits in the magnetic field is considered only for as long as it establishes the lower value of the magnitude square of the scattered momentum as previously discussed.

\textbf{The Weak Localization Magnetoconductivity -} Since the experimental detection of the WL occurs at magnetic fields of the order $10^{-4}$T, the quantization in Landau levels of the electron states introduces differences between consecutive $q$ values of the order $2eB/\hbar \simeq 10^{-2}q_{max}$, where $q_{max} = 1/\sqrt{D\tau_1}$. Consequently, we consider that $q$ varies quasi-continuously as in the absence of the magnetic field, thus allowing the simultaneous consideration of $q_x$ and $q_y$ as quasi-continuous variables, with the important caveat that the minimum value of $q$ attainable is imposed by the quantization condition, which for $n = 0$ establishes $q_{\min}^2 = 2eB/\hbar$.

The magnetoconductivity is calculated by integrating in the $\mb {q}$ space the inverse of the eigenvalues obtained in Eqs.~ (\ref{eq:singlet}), (\ref{eq:mode2}) and (\ref{eq:dispersion}), with input from (\ref{eq:taueff}). The origin of the integral is chosen, in each case, at the point where the dispersion equations reach a minimum, with the magnitude of the scattering momentum  between $q_{\min} = \sqrt{2eB/\hbar}$ and $q_{\max} = 1/\sqrt{D\tau_1}$. The logarithmic terms thus generated are then averaged over the angular distribution, defined in the usual way as $\int_0^{2\pi} d\varphi/2\pi$,
\begin{align}
&\Delta \sigma (B) = -\frac{e^2}{4\pi^2\hbar}\left\{-\ln\left(\frac{\tau_1^{-1}+\tau_\phi^{-1}}{\frac{2DeB}{\hbar}+\tau_\phi^{-1}}\right)\right.\nonumber\\
&\left.+  \ln\left[\frac{\tau_1^{-1}+\tau_{\phi}^{-1}+D({Q}_A^2 + Q_3^2)}{\frac{2DeB}{\hbar}+ \tau_\phi^{-1}+ D({Q}_A^2 + Q_3^2)}\right]\right.\nonumber\\
&\left.+2\left\langle\ln\left[\frac{\tau_1^{-1} + \tau_\phi^{-1}+\tau_{eff}^{-1}(\varphi)}{\frac{2DeB}{\hbar} +\tau_\varphi^{-1}+ \tau_{eff}^{-1}(\varphi}\right]\right\rangle_{\mbox{av}}\right\}\;.\label{eq:final}
\end{align}

\begin{figure}
{\includegraphics[width=\columnwidth]{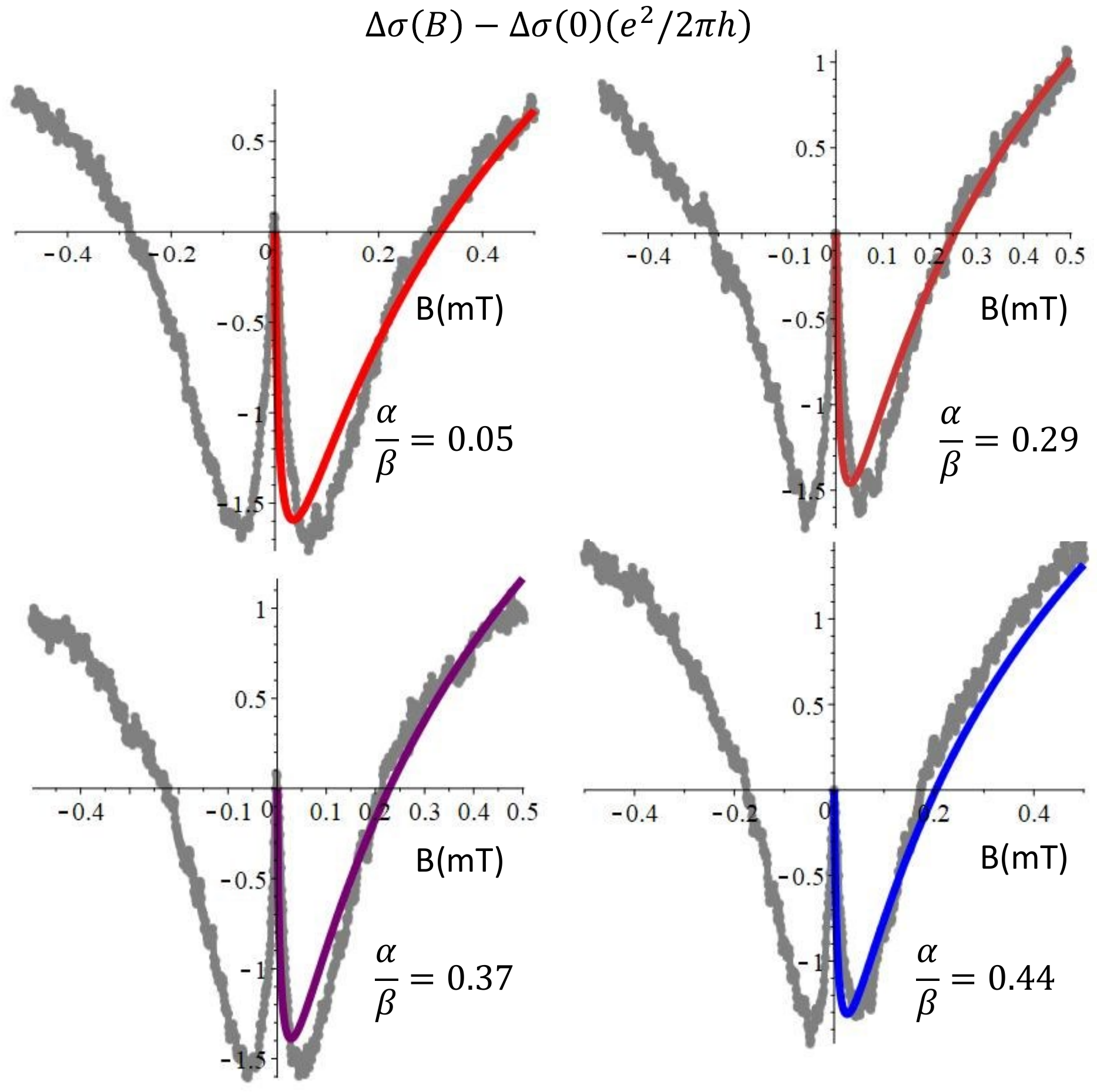}}
\caption{Experimentally determined magnetoconductivity corrections $\Delta \sigma (B) - \Delta \sigma (0)$, in $e^2/2\pi \hbar$ units, plotted as a function of the magnetic field $B$ (mT), are compared with the theoretical result in Eq.~(\ref{eq:final}) (continuous line) for $\alpha/\beta = 0.05, 0.29, 0.37, 0.44$.}
\label{test}
\end{figure}

\begin{figure}
    {\includegraphics[width = \columnwidth]{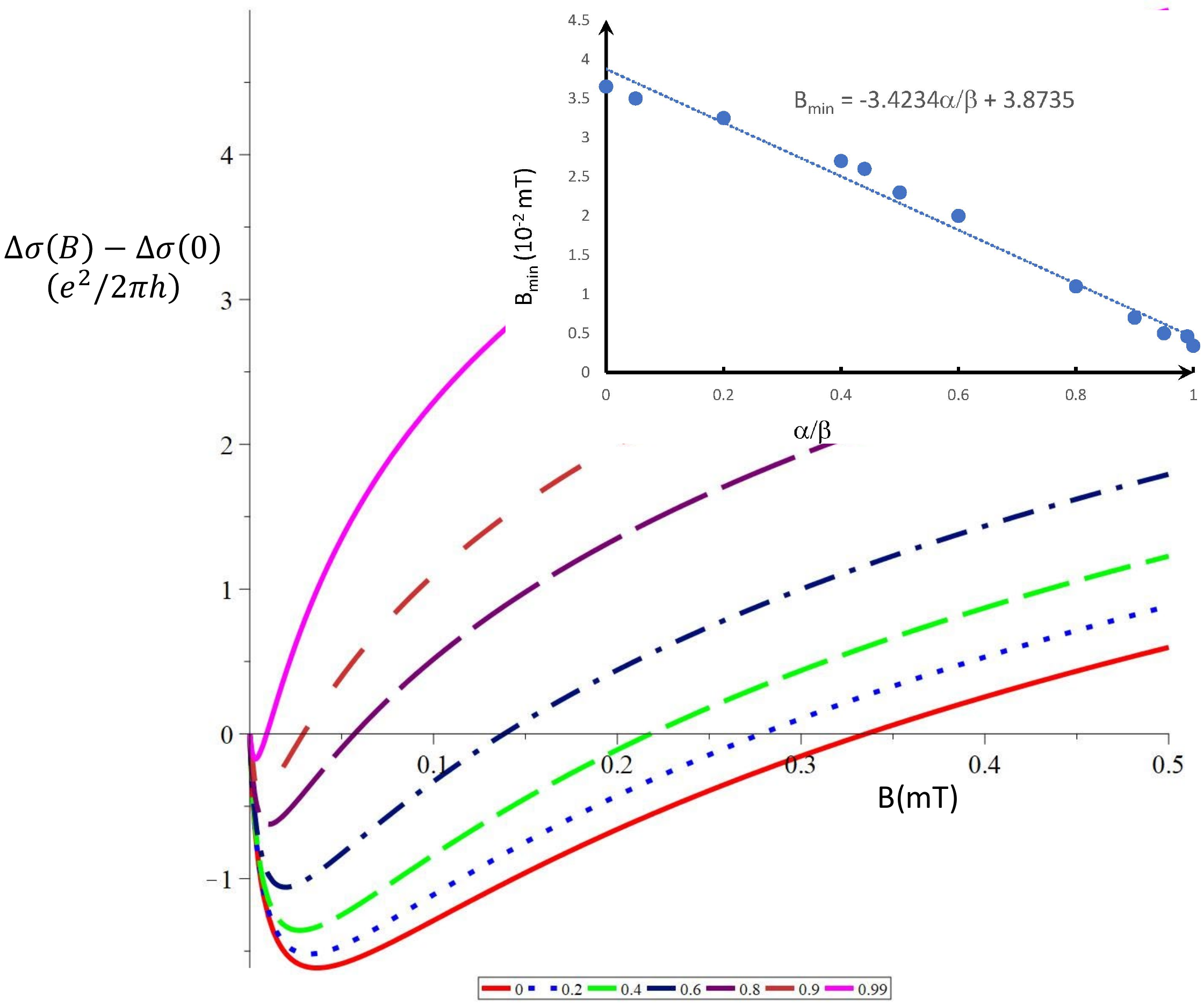}}
 \caption{\small Magnetoconductivity corrections are plotted as a function of the ratio of the linear SOI couplings, $\alpha/\beta$. In the inset, we track the magnetic field value $B_{\min}$ at which the magnetoconductivity reaches a minimum as a function of $\alpha/\beta$.}
 \label{fig2}
\end{figure}

In the limit $\alpha = 0$ (or $\beta = 0$), when $Q_S = Q_A$, the separation in the momentum space of the triplet Cooperon modes $1$ and $3$ is $2\mb q_0/\cos\theta = 2Q_S(\cos \varphi, \sin \varphi)$ parallel with $\mb q$ and $\tau_{eff}^{-1} = D(Q_S^2 + 3Q_3^2)/2$. Within the limits of our approximation, this solution is {\it{exact}}. When $B \rightarrow 0$, this expression is markedly different from the corresponding equation in Ref.~\cite{iordanskii} which considers only the diagonal elements of the triplet Cooperon eigenvalue matrix.

 When $\alpha \rightarrow \beta$, $Q_A\ll Q_S$, $\tau_{eff}^{-1} \simeq (Q_A^2 + 3Q_3^2)/2$, and $\mb q_0/\cos\theta \simeq \{Q_S, 0\}$ becomes almost parallel with the $\hat{x}$ axis. The separation in the momentum space of $2Q_s\hat{x}$ describes the formation of the two Fermi populations spin-polarized in opposite directions by the effective magnetic field associated with the SO coupling along the in-plane $\hat{z}$ known to lead, when $Q_3 = 0$ to the persistent spin helix state described in Ref.~\cite{bernevig}.
In this limit, the localization term determined by the spin relaxation along the $\hat{z}$ axis approaches in value the antilocalization correction, with an exact cancelation at $\alpha = \beta$ and $\beta_3 = 0$. For the same parameters, Eq.~(\ref{eq:final}) generates numerical values that are within $1\%$ from the fully quantized version of the magnetoconductivity derived in  Ref.~\cite{prx}.

In Fig.~\ref{test}, numerical estimates of $\Delta \sigma(B)$ given by Eq.~(\ref{eq:final}), relative to its value in the absence of the magnetic field $\Delta \sigma (0)$, are plotted against four sets of data obtained for different values of the $\alpha/\beta$ ratio in a GaAs quantum well with the following paratemeters: electron density $n = 7\times 10^{15}$ m$^{-2}$, experimentally determined transport times $\tau_1 = 2.06\times 10^{-12}$ s, $\tau_3 = \tau_1/3$, dephasing time $\tau_\phi = 10^{-9}$ s, cubic Dresselhaus coefficient, $\gamma = 12.6$ eV{\AA}, $\langle k_z^2\rangle = 3.9 \times 10^{16}$ m$^{-2}$ \cite{prx-1}.
We focus on the low and intermediate $\alpha/\beta$ values because no analytic expressions are presently available in this regime, where we find an excellent qualitative and quantitative agreement with the data.

The result of Eq.~(\ref{eq:final}) is plotted for the whole range of $\alpha/\beta$ values in Fig.~\ref{fig2}. The displacement of the magneto-conductivity minimum toward lower field values as $\alpha/\beta$ approaches $1$, considered a trademark of the SO coupled systems, is displayed in the inset. A fit to the numerical results, indicate that the field value for which the minimum is reached decreases quasi-linearly with $\alpha/\beta$. Obviously, the numerical parameters of the regression line are system dependent.

The numerical values of Eq.~(\ref{eq:final}) in the $\alpha \rightarrow \beta$ limit are larger than those discussed in Ref.~\cite{kammermeier}. The difference is a result of the magnetic field being introduced here only through the lower limit of the integral over $q$ which contributes a term $Dq_{\min}^2 = 2DeB/\hbar$ to the denominator in the final expression. Although this is identical to the magnetic shift rate in Refs.~\cite{ketterman, kammermeier}, there $1/\tau_B = 2DeB/\hbar$ is considered as a supplementary dephasing rate for the whole range of the momentum integral leading to a significant suppression of the magnetoconductivity value.

{\textbf{Conclusion -}} The qualitative and quantitative fit of the data with the analytical expression for the magnetoconductivity obtained in this algorithm suggest that, in evaluating the WL corrections in a SO coupled system placed in a perpendicular magnetic field, the anisotropy of the spin relaxation rates along the low spin-symmetry axes, here $\hat{y}$ and $\hat{x}$, plays a fundamental role. The mode associated with the high spin symmetry axis of the system, that along which in the limit $\alpha = \beta$ the spin projection becomes a quantum number, $\hat{z}$ here, compensates the antilocalization contribution. These results hold for all values of the SO constants. The closed-form expression obtained here provides a straightforward tool for the SO parameter extraction from experimental transport measurements.

\textbf{Acknowledgements -} This work was supported by the Swiss Nanoscience Institute (SNI), NCCR QSIT, Swiss NSF, European Microkelvin Platform EMP (DMZ), Brazilian grants FAPESP (SPRINT program), CNPq, PRP/USP (Q-NANO).

\end{document}